\documentclass[aps,prd,reprint,amsmath,amssymb,superscriptaddress]{revtex4-1}
\usepackage{bm}
\usepackage{graphicx}
\usepackage[colorlinks]{hyperref}
\usepackage{mathrsfs}
\usepackage{times}

\let\de=\partial
\newcommand\dd{\text{d}}
\newcommand\im{\text{i}}
\newcommand\De{\mathscr{D}}
\newcommand\La{\mathscr{L}}
\newcommand\Ha{\mathscr{H}}
\newcommand\Aa{\mathcal{A}}
\newcommand\hd{\hat d}
\newcommand\hdb{\hat{\vek d}}
\newcommand\hlb{\hat{\vek l}}
\newcommand\kf{k_\text{F}}
\newcommand\cs{c_\text{s}}
\newcommand\gr[1]{\mathrm{#1}}
\newcommand\vek[1]{\bm{#1}}
\DeclareMathOperator{\Tr}{Tr}

\begin{document}

\title{Helical spin texture in a thin film of superfluid $^3$He}

\author{Tom\'{a}\v{s} Brauner}
\email{tomas.brauner@uis.no}
\affiliation{Department of Mathematics and Physics, University of Stavanger, 4036 Stavanger, Norway}

\author{Sergej Moroz}
\email{sergej.moroz@tum.de}
\affiliation{Department of Physics, Technical University of Munich, 85748 Garching, Germany}

\begin{abstract}
We consider a thin film of superfluid $^3$He under conditions that stabilize the A-phase. We show that in the presence of a uniform superflow and an external magnetic field perpendicular to the film, the spin degrees of freedom develop a nonuniform, helical texture. Our prediction is robust and relies solely on Galilei invariance and other symmetries of $^3$He, which induce a coupling of the orbital and spin degrees of freedom. The length scale of the helical order can be tuned by varying the velocity of the superflow and the magnetic field, and may be in reach of near-future experiments.
\end{abstract}

\maketitle


\section{Introduction}

The experimental discovery of superfluidity in $^3$He~\cite{Osheroff1972,*Osheroff1972b} was a major breakthrough in low-temperature physics. The unconventional pairing of fermions in this system provided one of the first examples of topological quantum matter. The intricate symmetry-breaking patterns realized in $^3$He give rise to a number of unexpected phenomena which have kept both theorists and experimentalists busy for nearly a half century~\cite{vollhardt,Leggett:1975rmp,*volovik1992exotic,*volovikbook,*Mizushima:2016mi,Kleinert:2017kl}.

Recent advances in nanofabrication made it possible to study superfluidity experimentally under well-controlled conditions in $^3$He confined to two spatial dimensions~\cite{Levitin:PhD,Levitin2013,*Levitin:NatComm}. Two-dimensional confinement leads to a substantial modification of the phase diagram of superfluid $^3$He. In particular, at zero temperature it is the chiral A-phase that is energetically stabilized in a film with a thickness of the order of a few times the superfluid coherence length $\xi_0$~\cite{Vorontsov2003}.

\begin{figure}[t]
\begin{center}
\includegraphics[width=\columnwidth]{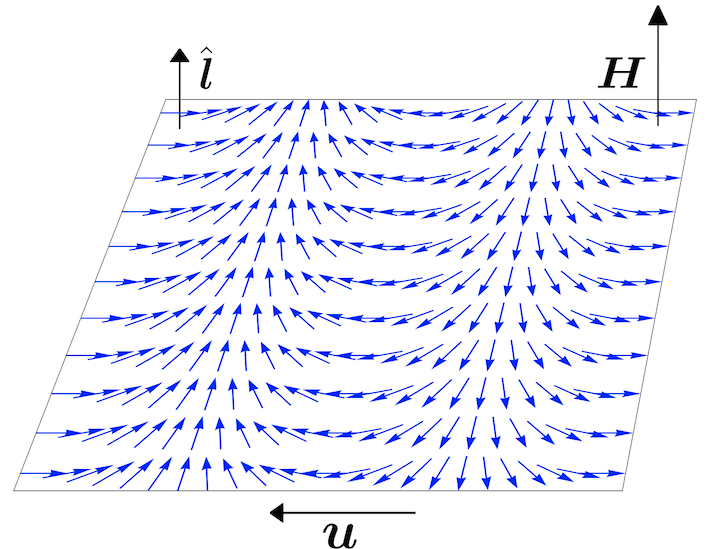}
\caption{Helical spin texture in a film of $^3$He-A. The orbital vector $\hlb$ is forced by surface interactions to be perpendicular to the film. The magnetic field $\vek H$ is \emph{chosen} to point in the same direction. The local, in-plane spin vector $\hdb$ varies along the superflow velocity $\vek u$, but remains uniform in the transverse direction.}
\label{fig1}
\end{center}
\end{figure}

Motivated by these developments, we analyze in this paper the low-energy spin physics in the A-phase of quasi-two-dimensional $^3$He at zero temperature. We use the effective field theory approach, based solely on symmetry and the low-energy degrees of freedom. Our main result is that the presence of a uniform superflow and a magnetic field $H\gtrsim30\text{ G}$, perpendicular to the $^3$He film, induces a nonuniform, planar helical texture (see Fig.~\ref{fig1}) in the ground state of the spin degrees of freedom. The pitch of the helical texture depends, apart from the macroscopic superflow velocity and the magnetic field, on a sole intrinsic observable: the phase velocity of spin waves. The pitch can be tuned by varying the former two macroscopic parameters, and within near-future experiments with superfluid $^3$He films, it may reach the centimeter range.

Owing to the rich structure of the order parameter, the precise form of the ground state of superfluid $^3$He usually depends on many factors, including geometrical constraints (boundary conditions), interaction with external fields, and last but not least, the weak dipole (spin-orbit) coupling between the spin and orbital degrees of freedom. This results in a large number of possible textures in superfluid $^3$He, depending on precise external conditions~\cite{vollhardt}. Thus, for instance, similar helical textures were previously predicted in bulk $^3$He~\cite{LinLiu:1978} and in $^3$He confined to a nanotube~\cite{Wiman:2018sa}. Likewise, a periodic texture was predicted for the A-phase of $^3$He confined to a thin slab~\cite{Hu:1979PRL}. The textures proposed in Refs.~\cite{LinLiu:1978,Hu:1979PRL} depend crucially on the presence of the dipole interaction.

The texture found in this paper is fundamentally different in that it does not rely on the presence of the dipole interaction. In contrast, it is a robust consequence of Galilei invariance and other symmetries of $^3$He. The only assumptions we make, that set constraints on possible experimental realization of this novel texture, are: (i) a slab geometry that stabilizes the A-phase, and (ii) a magnetic field strong enough to rotate the spin vector $\hdb$ into the slab plane.

The paper starts in Sec.~\ref{sec:2d3He} with an overview of the essentials of quasi-two-dimensional ${}^3$He, including its symmetries and some basic order-of-magnitude estimates, relevant for its experimental realization. In Sec.~\ref{sec:EFT}, we then develop the low-energy effective field theory of spin in the A-phase, stressing the role of Galilei invariance. This is followed by a detailed derivation of the helical texture in the ground state in Sec.~\ref{sec:groundstate}. In Sec.~\ref{sec:spectrum}, we discuss the excitation spectrum above the helical texture, and its possible signatures through nuclear magnetic resonance (NMR) spectroscopy. Some further technical details are relegated to two appendices.


\section{Quasi-two-dimensional ${}^3$H\lowercase{e}}
\label{sec:2d3He}

Bulk ${}^3$He at zero temperature and low pressures features the isotropic B-phase. The ground state, however, changes when ${}^3$He is confined to a narrow slab. Weak-coupling theory predicts~\cite{Vorontsov:2007Sa} that the A-phase is stabilized for slab thickness $D\lesssim9\xi_0$, being separated from the B-phase by a stripe phase at $9\xi_0\lesssim D\lesssim13\xi_0$. While the question of the existence of the stripe phase remains unresolved by experiment, the stability of the A-phase in narrow slabs has been confirmed~\cite{Levitin:NatComm,Levitin2018}. Given that $\xi_0\approx70\text{ nm}$ for pressures below ca $2\text{ bar}$~\cite{Levitin:PhD}, and that the dipole interaction only becomes important at length scales above the order of $10\text{ $\mu$m}$~\cite{Kleinert:2017kl}, the latter will play a negligible role in our analysis.

The order parameter of the A-phase of ${}^3$He has the structure
\begin{equation}
\Delta_{ir}\propto\hd_i(\hat m_r+\im\hat n_r),
\label{OP}
\end{equation}
where $\hdb$ is a unit vector in the spin space and $\hat{\vek m},\hat{\vek n}$ are two orthogonal unit vectors in the orbital space. The three degrees of freedom contained in $\hat{\vek m},\hat{\vek n}$ can be encoded in a single vector, $\hlb\equiv\hat{\vek m}\times\hat{\vek n}$, and an overall complex phase $\theta$. Boundary effects induce an aligning force on $\hlb$ that tries to orient it perpendicularly to the surface. In the quasi-two-dimensional regime of ${}^3$He confined to a narrow slab, the $\hlb$-vector will be completely oriented to the direction normal to the slab, and the only active orbital degree of freedom will thus be the superfluid phase $\theta$. The total of three degrees of freedom, contained in $\hdb$ and $\theta$, correspond to the symmetry-breaking pattern in the A-phase in two spatial dimensions~\cite{vollhardt},
\begin{equation}
\gr{SU(2)_S\times SO(2)_L\times U(1)_\phi}\to\gr{U(1)_S\times U(1)_{\phi-L}},
\label{SSB}
\end{equation}
where ``S'' and ``L'' refer respectively to spin and orbital symmetries and $\gr{U(1)_\phi}$ stands for the particle number symmetry.

The dipole interaction breaks the independent spin and orbital symmetries down to the diagonal $\gr{SO(2)_{L+S}}$ subgroup. In the absence of other symmetry-breaking perturbations, it aligns the $\hdb$-vector (anti)parallel to $\hlb$. To overcome this weak aligning force and make the spin vector $\hdb$ oriented in the slab plane, we assume the presence of a magnetic field $\vek H$, perpendicular to the slab. The desired orientation of the $\hdb$-vector will be achieved provided $\vek H$ is stronger than the characteristic field of the dipole interaction, $H_\text{d}\approx 30\text{ G}$~\cite{vollhardt}. On the other hand, the magnetic field should not be too strong so as not to distort significantly the order parameter. Taking the temperature scale of the order parameter as $T_\Delta\sim1\text{ mK}$, we can estimate the corresponding critical angular frequency as $k_\text{B}T_\Delta/\hbar\sim100\text{ MHz}$. Current experiments typically operate at Larmor frequencies of spin precession $f_\text{L}=\omega_\text{L}/(2\pi)\sim1\text{ MHz}$, corresponding to magnetic field $H\approx300\text{ G}$~\cite{Levitin:PhD}. This satisfies with a good margin both bounds.

Finally, recall that the superfluid becomes unstable when the superflow velocity $\vek u$ exceeds the Landau critical velocity. For the A-phase of ${}^3$He, this is of the order of $u_\text{cr}\approx5\text{ cm/s}$~\cite{Kleinert:2017kl}. The superflow velocity in actual experiments on ${}^3$He films is typically much lower, in the sub-$\text{mm/s}$ range~\cite{Sachrajda1985,*Davis1988}.


\section{Low-energy effective theory}
\label{sec:EFT}

The dynamics of A-phase of quasi-two-dimensional ${}^3$He at low energies is dominated by the soft degrees of freedom corresponding to the symmetry-breaking pattern~\eqref{SSB}, that is, the variables $\hdb$ and $\theta$. In this paper, we assume that the superflow, defined by its velocity $\vek u=\vek\nabla\theta/m$, constitutes a fixed background for the dynamics of the spin vector $\hdb$. This is a reasonable assumption for $u\ll u_\text{cr}$, and can be justified formally using the power counting of the low-energy effective theory~\cite{Son2006,*Fujii2016}. With this assumption, the low-energy dynamics of the order parameter $\hdb$ can be fully captured by an effective theory for $\hdb$ alone.

The effective theory must respect all the symmetries of the microscopic interactions among ${}^3$He atoms. The spacetime symmetries include space and time translations, Galilei invariance, spatial rotation invariance $\gr{SO(2)_L}$, two-dimensional parity $P$ (under which $x\leftrightarrow y$) and time reversal $T$. The internal symmetries include the spin rotation invariance $\gr{SU(2)_S}$ and the particle number symmetry $\gr{U(1)_\phi}$.

Under an infinitesimal boost, $\vek x'=\vek x+\vek vt$, the superfluid phase $\theta$ shifts as $\theta'(\vek x')=\theta(\vek x)+m\vek v\cdot\vek x$. Galilei invariance then requires that time derivatives of other, boost-invariant fields only enter the action through the ``material derivative'', $\tilde\de_t\equiv\de_t+\vek u\cdot\vek\nabla$. To the leading order in the derivative expansion, the effective spin Lagrangian density then reads~\footnote{In spin systems, it is customary to give the Lagrangian a prefactor called spin stiffness. Such an overall scale has no bearing on our analysis and is therefore dropped. The same remark applies to the ensuing Hamiltonian~\eqref{ham2} and spin current~\eqref{spincurrent}.}
\begin{equation}
\La=\frac12(D_t\hdb+u_rD_r\hdb)^2-\frac{\cs^2}2(D_r\hdb)^2+\La_\text{dip}.
\label{lag}
\end{equation}
Here $\cs$ is the phase velocity of spin waves in the absence of background fields. The covariant derivative of the $\hdb$-vector is defined by
\begin{equation}
D_\mu\hdb\equiv\de_\mu\hdb+\vek A_\mu\times\hdb,
\label{covder}
\end{equation}
where $\vek A_\mu$ is the gauge field of the $\gr{SU(2)_S}$ group. In presence of a magnetic field $\vek H$ and no other external fields, it reads $\vek A_\mu=\delta_{\mu t}\gamma\vek H$~\cite{Frohlich:1993gs}, where $\gamma\approx-20379\text{ (G$\cdot$s)}^{-1}$ is the gyromagnetic ratio of the ${}^3$He nucleus~\cite{Wheatley1975}. Finally, the symmetry-breaking perturbation $\La_\text{dip}$ represents the dipole interaction,
\begin{equation}
\La_\text{dip}=\frac12\Omega_\text{L}^2(\hlb\cdot\hdb)^2,
\label{dip}
\end{equation}
where $\Omega_\text{L}$ is the so-called Leggett frequency, corresponding to the dipole field $H_\text{d}$. We stress that the coupling to the magnetic field, defined by Eq.~\eqref{covder}, is \emph{not} a perturbation in the same sense as the dipole coupling. Namely, it is completely fixed by the $\gr{SU(2)_S}$ invariance, and involves no new, a priori arbitrary, parameters.

In two spatial dimensions, the term $\epsilon^{rs}\hdb\cdot(D_r\hdb\times D_s\hdb)$ is also consistent with the continuous symmetries of the system. This term is, however, prohibited by the discrete parity and time-reversal symmetries.

Our construction above is completely general and relies on the symmetries of the system only. Given a microscopic model of a thin film of $^3$He, on can alternatively derive the effective Lagrangian~\eqref{lag} by integrating out the fermionic degrees of freedom. Such an approach allows one to fix the spin wave velocity in terms of the parameters of the microscopic model. To complement our general construction presented here, we perform this calculation for the Bogoliubov-de-Gennes mean-field theory in Appendix~\ref{app:micro}.


\section{Ground state texture}
\label{sec:groundstate}

We are now interested in the ground state of the system in the presence of a uniform background superflow and an external magnetic field perpendicular to the film, see Fig. \ref{fig1}. To that end, we first  compute the canonical Hamiltonian density, $\Ha=\de_t\hdb\cdot\de\La/\de(\de_t\hdb)-\La$,
\begin{equation}
\Ha=\frac12(\de_t\hdb)^2-\frac12(\gamma\vek H\times\hdb+u_r\de_r\hdb)^2+\frac{\cs^2}2(\de_r\hdb)^2-\frac12\Omega_\text{L}^2(\hlb\cdot\hdb)^2.
\label{ham2}
\end{equation}
Given the way the temporal derivatives enter the Hamiltonian, the ground state will obviously be time-independent. The Ha\-mil\-to\-nian for static field configurations can then be cast as
\begin{equation}
\begin{split}
\Ha=&-\frac{\gamma^2}2(\vek H\times\hdb)^2-\frac12\Omega_\text{L}^2(\hlb\cdot\hdb)^2\\
&-\gamma u\de_x\hdb\cdot(\vek H\times\hdb)+\frac{c_\text{eff}^2}2(\de_x\hdb)^2+\frac{\cs^2}2(\de_y\hdb)^2,
\end{split}
\end{equation}
where, without loss of generality, we chose the $x$-axis along the superflow. We also defined $c_\text{eff}^2\equiv\cs^2-u^2$. Note that in practice, the Landau critical velocity $u_\text{cr}$ is much smaller than the spin-wave velocity $\cs$, hence the coefficient $c_\text{eff}^2$ is always positive and approximately equal to $\cs^2$. Next, we combine the terms containing $\de_x\hdb$ and rewrite $(\vek H\times\hdb)^2=\vek H^2-(\vek H\cdot\hdb)^2$, which leads to
\begin{equation}
\begin{split}
\Ha={}&\Ha_0+\frac{\gamma^2}2\left(1+\frac{u^2}{c_\text{eff}^2}\right)(\vek H\cdot\hdb)^2-\frac12\Omega_\text{L}^2(\hlb\cdot\hdb)^2\\
&+\frac{c_\text{eff}^2}2\left(\de_x\hdb-\frac{\gamma u}{c_\text{eff}^2}\vek H\times\hdb\right)^2+\frac{\cs^2}2(\de_y\hdb)^2,
\end{split}
\label{minHam}
\end{equation}
where $\Ha_0\equiv-\frac{\gamma^2}2\bigl(1+\frac{u^2}{c_\text{eff}^2}\bigr)\vek H^2$. This makes it clear that for $\vek H\parallel\hlb$ and $H>H_\text{d}$, or equivalently $|\gamma|H>\Omega_\text{L}$, the following conditions must be satisfied in the state of lowest energy,
\begin{equation}
\vek H\cdot\hdb=0,\qquad
\de_x\hdb=\frac{\gamma u}{c_\text{eff}^2}\vek H\times\hdb,\qquad
\de_y\hdb=\vek 0.
\label{larmor}
\end{equation}
The unique solution up to an overall spin rotation is given by in-plane Larmor precession of the $\hdb$-vector with the coordinate $x$ along the superflow playing the role of time, see Fig.~\ref{fig1}.

The pitch of the helical texture follows from Eq.~\eqref{larmor} and can be expressed in terms of easily measurable quantities as
\begin{equation}
\lambda=\frac{1}{f_\text{L}}\frac{\cs^2-u^2}u=\frac{2\pi}{|\gamma|H}\frac{\cs^2-u^2}u.
\label{wavelength}
\end{equation}
Assuming that $u\ll\cs$ and approximating the spin-wave velocity by $\cs\approx20\text{ m/s}$~\cite{3Helibrary}, we get a numerical estimate for the pitch in terms of the tunable parameters $f_\text{L}$, or $H$, and $u$,
\begin{equation}
\lambda\approx40\text{ cm}\times\left(\frac{f_\text{L}}{\text{MHz}}\,\frac{u}{\text{mm/s}}\right)^{-1}\approx120\text{ m}\times\left(\frac{H}{\text{G}}\,\frac{u}{\text{mm/s}}\right)^{-1}.
\end{equation}
Taking $f_\text{L}\approx1\text{ MHz}$ and $u\approx1\text{ mm/s}$ as typical for current experiments [see the discussion below Eq.~\eqref{SSB}] gives $\lambda\approx40\text{ cm}$. Since the size of the experimental cell in current experiments lies in the centimeter range~\cite{Levitin:PhD}, either $f_\text{L}$ or $u$ (or both) has to be increased moderately for the helical texture to be directly observable. The latter might be achieved by using a different geometry, either by studying a thin layer of ${}^3$He under rotation, or by using oscillating superflow~\cite{PhysRevB.91.024503}. Even if the whole pitch turns out to be too long, it should still be possible to observe the effect through chirality of spin-spin correlations.

Let us now mention some theoretical aspects of the discovered helical texture. First of all, the derivation of the ground state was carried out in a fixed reference frame attached to the slab confining the ${}^3$He sample; the parameter $\vek u$ measures the velocity of the superflow with respect to the slab. The same result can, however, be obtained in any other reference frame due to Galilei invariance; see Appendix~\ref{app:galilei} for details.

Second, the generation of dissipation-less spin currents has been of great theoretical as well as practical interest lately (see e.g.~Ref.~\cite{Zutic:RMP,*Linder:NatPhy}), and the structure of the helical ground state might suggest that it carries such a current. The Noether current of the $\gr{SU(2)_S}$ spin symmetry reads
\begin{equation}
\vek j^\mu=\hdb\times\frac{\de\La}{\de(\de_\mu\hdb)},
\label{spincurrent}
\end{equation}
and for the spin texture Eq.~\eqref{larmor} only has a temporal component, $\gamma\vek H\cs^2/c_\text{eff}^2$. The spin current is therefore zero in the reference frame used here, but due to the nonzero spin density, it will be nonzero in any other inertial reference frame, see Appendix~\ref{app:galilei}.

Third, previous theoretical work~\cite{Volovik:1989yak,*Stone2004} discovered that the effective theory of spin in a superfluid $^3$He-A film contains a topological Hopf term, responsible for the quantum statistics of skyrmions and quantized spin Hall effect. The Hopf term is defined by the Lagrangian
\begin{equation}
\La_\text{Hopf}=\frac1{32\pi^2}\int\dd^2\vek x\,\dd t\,\epsilon^{\mu\nu\lambda}\mathscr A_\mu\mathscr F_{\nu\lambda},
\end{equation}
where $\mathscr F_{\mu\nu}\equiv\de_\mu\mathscr A_\nu-\de_\nu \mathscr A_\mu\equiv\hdb\cdot(\de_\mu\hdb\times\de_\nu\hdb)$ is an auxiliary composite gauge field. The Hopf term was not included in our effective theory, being formally of higher order in the derivative expansion. Moreover, our helical texture only varies in one spatial direction, hence it carries zero skyrmion number and the Hopf term accordingly vanishes.

Fourth, the ground state can be found using the Hamiltonian~\eqref{minHam} also for other orientations of the magnetic field than perpendicular to the slab. In the ideal limit of exact spin symmetry, $\Omega_\text{L}\to0$, the ground state will correspond to an analogous helical texture featuring precession of the $\hdb$-vector around the $\vek H$-vector. A nonzero dipole coupling will in general lead to a distortion of the helix when $\vek H\nparallel\hlb$.

Finally, note that the helical texture can also be derived using Ginzburg-Landau theory, which is based on power expansion in both the order parameter and its derivatives. To that end one can use the energy functional for superfluid ${}^3$He in presence of an external magnetic field, derived in Ref.~\cite{Volovik1984}. Note, however, that the Ginzburg-Landau framework is only reliable close to the critical temperature for the superfluid phase transition. In contrast, our effective field theory setup is designed to work at zero temperature and is organized as an expansion in derivatives of the order parameter fluctuations.


\section{Excitation spectrum}
\label{sec:spectrum}

The basic tool for identification of nonuniform textures in ${}^3$He is nuclear magnetic resonance (NMR) spectroscopy~\cite{Leggett:1973PRL}. To understand possible NMR signatures of our helical texture, we need to determine the excitation spectrum. To that end, we write the $\hdb$-vector in the ground state as
\begin{equation}
\langle\hd_1\rangle=\cos\alpha x,\qquad
\langle\hd_2\rangle=\sin\alpha x,\qquad
\langle\hd_3\rangle=0,
\end{equation}
where $\alpha\equiv\gamma uH/c_\text{eff}^2$. Next, introduce the ``comoving'' spin variable $\hdb'$ through
\begin{equation}
\hdb(\vek r)=\begin{pmatrix}
\cos\alpha x & -\sin\alpha x & 0\\
\sin\alpha x & \cos\alpha x & 0\\
0 & 0 & 1
\end{pmatrix}\hdb'(\vek r),
\end{equation}
in which the ground state is trivial, $\langle\hdb'\rangle=(1,0,0)$. Upon this redefinition, the Lagrangian~\eqref{lag} becomes, up to a constant,
\begin{equation}
\begin{split}
\La={}&\frac12(\de_t\hdb'+u\de_x\hdb')^2-\frac12\cs^2(\de_r\hdb')^2\\
&+\gamma H\left(1+\frac{u^2}{c_\text{eff}^2}\right)(\hd'_1\de_t\hd'_2-\hd'_2\de_t\hd'_1)\\
&-\frac12\left[\gamma^2H^2\left(1+\frac{u^2}{c_\text{eff}^2}\right)-\Omega_\text{L}^2\right]\hd'^2_3.
\end{split}
\label{Lbilin}
\end{equation}
Since the ground state is oriented in the $\hd'_1$ direction, the spectrum is determined by the part of the Lagrangian bilinear in $\hd'_{2,3}$. The dispersion relations of the two modes, corresponding to $\hd'_{2,3}$, can be read off the first and third line of Eq.~\eqref{Lbilin},
\begin{equation}
\omega_{2,3}(\vek k)=uk_x+\sqrt{\cs^2\vek k^2+\mu_{2,3}^2},
\label{disp}
\end{equation}
where
\begin{equation}
\mu_2=0,\qquad
\mu_3=\sqrt{\gamma^2H^2\left(1+\frac{u^2}{c_\text{eff}^2}\right)-\Omega_\text{L}^2}.
\end{equation}
Note that $\hd'_2$ remains gapless in spite of the presence of the external magnetic field and the dipole coupling. This reflects the exact $\gr{U(1)_S}$ symmetry corresponding to in-plane spin rotations, which is spontaneously broken in the ground state.

In the theory of NMR response due to Leggett~\cite{vollhardt,Leggett:1973PRL}, the resonance frequencies are obtained by solving the equations of motion for the $\hdb$-vector and the operator of total spin. At zero temperature, where our effective theory setup applies, the equation of motion for spin is a consequence of that for the $\hdb$-vector though.

The frequency of collective spin oscillations, probed by NMR with a uniform magnetic field, corresponds to the spin-wave dispersion relation~\eqref{disp} at $\vek k=\vek0$, and is given by $\mu_{2,3}$. The tiny $u$-dependent shift of the resonance frequency of the $\hd_3'$ mode can in principle be used as evidence for our helical texture. The relative shift of the frequency is essentially independent of the magnetic field and for superflow in the $\text{mm/s}$ range is of the order $u^2/(2c^2_{\text{eff}})\sim 10^{-9}$, which is at the frontier of resolution in current NMR experiments.


\section{Conclusions}
\label{sec:conclusions}

Galilei invariance is known to impose powerful constraints on effective theories of nonrelativistic superfluids~\cite{Son2006,*Fujii2016,Greiter:1989qb,*Andersen2002,*Hoyos2013,*Moroz2014a}. In this paper we argued that in case of a thin film of $^3$He-A, it inevitably leads to a coupling between superflow and spin degrees of freedom, an effect that could easily be overlooked by considering only the orbital and spin symmetries and their spontaneous breaking. Based on this observation, we predicted that the ground state of a superfluid film of $^3$He-A in presence of a uniform superflow and an external magnetic field perpendicular to the film features a nonuniform, helical texture. The helix pitch depends only on the phase velocity of spin waves, the superflow velocity and the magnetic field, and can be tuned by varying the latter two.

In order to gain a better grasp on the phenomenological implications of our prediction, it would be desirable to study the effects of nonzero temperature. On the one hand, this would help to clarify in what temperature range the helical texture represents the equilibrium state of a thin film of superfluid $^3$He-A. By the same token, it would be important to understand the role of thermal fluctuations in the equilibrium state.

In view of experimental prospects for detection of the predicted texture, it would likewise be interesting to extend our study of spin physics of two-dimensional $^3$He-A superflow to other geometries, including rotating and oscillating superflow under external magnetic field.

Finally, given the model-independent nature of the effective theory used here, it would be interesting to search for other systems where the combination of uniform external fields and Galilei invariance might lead to a nonuniform ground state.


\begin{acknowledgments}
We would like to thank Lev Levitin, David Schmoranzer and Grigory Volovik for illuminating discussions and Peter W\"olfle for useful correspondence. The work of T.B.~is supported by the ToppForsk-UiS grant no.~PR-10614. The work of S.M.~is supported by the Emmy Noether Programme of German Research Foundation (DFG) under grant no.~MO 3013/1-1.
\end{acknowledgments}


\appendix

\section{Microscopic derivation of effective action}
\label{app:micro}

Here the effective theory for the spin and superfluid degrees of freedom will be derived from a microscopic fermionic model. To that end, we will first specify the microscopic theory and make sure that it has the desired symmetries. Subsequently, we will integrate out the fermionic degrees of freedom to obtain the effective action. For simplicity, the dipole interaction will be neglected here.


\subsection{Symmetries of ${}^3$He}

Both the low-energy effective theory and any microscopic model must respect the actual symmetries of ${}^3$He. In the three-dimensional bulk and in the absence of the dipole interaction, the total continuous global symmetry group of ${}^3$He is
\begin{equation}
G=\gr{SU(2)_S\times SO(3)_L\times U(1)_\phi},
\end{equation}
together with space and time translations and Galilei boosts. Here  $\gr{SU(2)_S}$ corresponds to spin rotations, $\gr{SO(3)_L}$ to spatial (orbital) rotations, and $\gr{U(1)_\phi}$ to the conservation of the number of Helium atoms (particle number). The order parameter of the A-phase as given in Eq.~(\ref{OP}) breaks the symmetry group $G$ spontaneously down to
\begin{equation}
H=\gr{U(1)_S\times U(1)_{\phi-L}}.
\end{equation}
The spin rotation group is broken to its $\gr{U(1)_S}$ subgroup by the $\hdb$-vector. The residual $\gr{U(1)_{\phi-L}}$ subgroup reflects the fact that the complex orbital vector $\hat{\vek m}+\im\hat{\vek n}$ is left invariant by a combination of a spatial rotation and a phase redefinition.

In the quasi-two-dimensional setup considered here, the rotation group $\gr{SO(3)_L}$ is reduced  to the $\gr{SO(2)_L}$ group of in-plane rotations. In the thin layer of ${}^3$He in the A-phase, the orbital $\hlb$-vector is aligned by boundary effects perpendicularly to the slab. As a consequence, the vectors $\hat{\vek m}$ and $\hat{\vek n}$ lie in the slab plane, and the residual $\gr{U(1)_{\phi-L}}$ symmetry is maintained. The low-energy degrees of freedom of the A-phase of quasi-two-dimensional ${}^3$He therefore follow from the symmetry-breaking pattern
\begin{equation}
\gr{SU(2)_S\times SO(2)_L\times U(1)_\phi}\to\gr{U(1)_S\times U(1)_{\phi-L}}.
\end{equation}


\subsection{Microscopic action}

We shall now consider an idealized theory of strictly two-dimensional ${}^3$He where the fermionic degrees of freedom are fully gapped in the A-phase. Without specifying a concrete microscopic interaction, we assume that the theory has been semi-bosonized. This leads to a Bogoliubov-de-Gennes-type theory that describes noninteracting fermions propagating on a background of collective pair fields. Following closely the notation introduced by Stone and Roy~\cite{Stone2004}, we write the \emph{Euclidean} Lagrangian of this microscopic mean-field theory as
\begin{equation}
\La=\frac12\Psi^\dagger(\de_\tau+\hat H)\Psi,\qquad
\hat H\equiv\begin{pmatrix}
\hat h & \hat\Delta\\
{\hat\Delta}^\dagger & -\hat h^T
\end{pmatrix},
\label{micL}
\end{equation}
where $\Psi\equiv\begin{pmatrix}\psi_\alpha &\psi^*_\alpha\end{pmatrix}^T$ is the Nambu spinor with $\alpha=\uparrow,\downarrow$. In addition,
\begin{equation}
\hat h\equiv-\frac1{2m}(\vek\nabla-\im\vek A-\im\vek B)^2-(A_0+B_0)
\end{equation}
is the one-particle Hamiltonian. It will turn out convenient to couple the microscopic fermionic theory to a set of background gauge fields for its internal symmetries. Thus, $A_\mu$ is the matrix-valued gauge field of the spin $\gr{SU(2)_S}$ group, whereas $B_\mu$ is the gauge field of the $\gr{U(1)_\phi}$ symmetry.

The physical content of Eq.~(\ref{micL}) can be highlighted by disposing of the Nambu notation and rewriting the Lagrangian, up to a surface term, as
\begin{equation}
\La=\psi^\dagger(\de_\tau+\hat h)\psi+\frac12(\psi^\dagger\hat\Delta\psi^*+\text{H.c.}).
\end{equation}
The pairing field $\hat\Delta$ must be antisymmetric as a consequence of the Pauli principle, and can be cast as
\begin{equation} \label{dus}
\hat\Delta=\frac{\Delta}{2\kf}\bigl(\hat P\hat\Sigma e^{\im\Phi}-e^{\im\Phi}\hat\Sigma\hat P^T\bigr),
\end{equation}
where
\begin{equation}
\begin{split}
\hat\Sigma&\equiv\im(\hdb\cdot\vek\sigma)\sigma_2,\\
\hat P&\equiv-\im(D_x+\im D_y).
\end{split}
\label{OP2}
\end{equation}
Here $\kf$ is the Fermi momentum, $\Delta$ the gap parameter, $\vek\sigma$ the vector of Pauli matrices, and the covariant derivatives with spatial and temporal indices are defined as
\begin{equation}
\begin{split}
\vek D&\equiv\vek\nabla-\im(\vek A+\vek B)\equiv\vek\nabla-\im\vek\Aa,\\
D_\tau&\equiv\de_\tau-(A_0+B_0)\equiv\de_\tau-\Aa_0.
\end{split}
\end{equation}
Finally, we used the shorthand notation $\Phi\equiv\frac2\hbar\theta$ for the collective field of the spontaneously broken $\gr{U(1)_\phi}$ symmetry. Note that our expression for $\hat\Delta$ differs somewhat from that of Stone and Roy~\cite{Stone2004}. The form (\ref{dus}) is necessary for maintaining the full gauge symmetry, as long as we wish to write the Lagrangian in terms of simple, covariant building blocks.

Let us now give explicit expressions for the symmetries of the Lagrangian. We will denote by $U$ a generic element of the $\gr{SU(2)_S\times U(1)_\phi}$ gauge group. It can be decomposed as $U=U_1U_2=U_2U_1$, using the natural notation for $U_1\in\gr{U(1)_\phi}$ and $U_2\in\gr{SU(2)_S}$. The transformation rules for the fermions and the gauge field $\Aa_\mu$ then read
\begin{equation}
\begin{split}
\psi&\to U\psi,\\
\vek\Aa&\to U\vek\Aa U^{-1}+\im U\vek\nabla U^{-1},\\
\Aa_0&\to U\Aa_0U^{-1}-U\de_\tau U^{-1}.
\end{split}
\label{gaugetransfo}
\end{equation}
The second and third line summarize the usual transformation rule for a non-Abelian gauge field, modified owing to the fact that we work in Euclidean space. The transformation rules for the collective fields $\hdb$ and $\Phi$ read accordingly
\begin{equation}
\begin{split}
\hdb\cdot\vek\sigma&\to U_2(\hdb\cdot\vek\sigma)U_2^{-1},\\
e^{\im\Phi}&\to U_1e^{\im\Phi}U_1=U_1^2e^{\im\Phi}=e^{\im\Phi}U_1^2.
\end{split}
\label{gaugetransfoa}
\end{equation}
The first line above implies
\begin{equation}
\hat\Sigma\to U_2\hat\Sigma U_2^T.
\label{Sig}
\end{equation}
Since the covariant derivatives transform by construction covariantly, $\hat P\to U\hat PU^{-1}$, one finds in the end that
\begin{equation}
\hat\Delta\to U\hat\Delta U^T.
\label{gaugetransfob}
\end{equation}

Based on Eqs.~(\ref{gaugetransfo}), (\ref{gaugetransfoa}) and (\ref{gaugetransfob}), we can conclude that the Lagrangian (\ref{micL}) is gauge-invariant under transformations from the $\gr{SU(2)_S\times U(1)_\phi}$ group as it should.


\subsection{Effective action}

By integrating out the fermions, we arrive at the effective action, given in Euclidean space by
\begin{equation}
\label{Seffdef}
S_\text{eff}=-\frac12\Tr\log(\de_\tau+\hat H)\equiv-\frac12\Tr\log\De^{-1}.
\end{equation}
This action is a functional of $\Phi$, $\hdb$ and $\Aa_\mu$, and inherits the gauge invariance of the microscopic action under a simultaneous gauge transformation of these fields. There is no anomaly involved in integrating out the fermions, since the symmetry transformation of the fermion field $\Psi$ is realized by a unitary similarity transformation of the Bogoliubov-de-Gennes (BdG) operator $\de_\tau+\hat H$, and thus does not affect its spectrum.

At this intermediate stage, it is convenient to use the gauge invariance of the effective action to remove the collective scalar fields. The variable $\hdb$ transforms in the vector, or adjoint, representation of $\gr{SU(2)_S}$ and can be rotated to any fixed direction by a local $\gr{SU(2)_S}$ transformation. In other words, there is a unitary matrix $V$ such that
\begin{equation}
\hdb\cdot\vek\sigma=V\sigma_2V^{-1},\qquad
\hat\Sigma=\im VV^T.
\label{groundstate}
\end{equation}
From Eqs.~(\ref{gaugetransfoa}) and (\ref{Sig}), we can see that both $\Phi$ and $\hdb$ can then be absorbed into a redefinition of the gauge field $\Aa_\mu$ by choosing
\begin{equation}
U_1=e^{-\im\Phi/2},\qquad
U_2=V^{-1}.
\label{U12}
\end{equation}
The effective action now depends solely on the composite gauge field, defined by Eq.~(\ref{gaugetransfo}) with the above choice for $U_{1,2}$. In the following, this composite gauge field will be denoted by the same symbol $\Aa_\mu$. Only at the very end of this section, we will restore the dependence of the action on the spin vector $\hdb$ and the phase $\Phi$.

To evaluate the effective action, we adopt a derivative expansion scheme. Since we are interested in the dynamics of \emph{small} fluctuations of the spin degrees of freedom, we shall count each derivative of $\hdb$ as order 1. At the same time, we allow for a finite \emph{uniform} velocity of the superflow background. Hence, one derivative acting on $\Phi$ will count as order 0, and every other derivative acting on the same field as order 1. As a consequence, the fields $A_\mu$ and $B_\mu$ are of order 1 and 0, respectively. We shall evaluate the effective action~(\ref{Seffdef}) to the leading order in both fields, which means order 2 for $A_\mu$ and order 0 for $B_\mu$. In this approximation, we can treat $\Aa_\mu$ as a constant fixed background. We need to expand to second order in $A_\mu$, whereas $B_\mu$ has to be resummed to all orders.

To facilitate the Taylor expansion in the non-Abelian gauge field $A_\mu$, it is suitable to split the BdG operator into parts of order zero, one and two in $A_\mu$,
$\De^{-1}=\De_0^{-1}+\De_1^{-1}+\De_2^{-1}$. Upon Fourier transforming to frequency $\omega$ and momentum $\vek p$,
\begin{align}
\notag
\De_0^{-1}&=\begin{pmatrix}
\im\omega+\frac{\vek\pi^2}{2m}-B_0 & \frac{\im\Delta p_+}{\kf}\\
-\frac{\im\Delta p_-}{\kf} & \im\omega-\frac{\tilde{\vek\pi}^2}{2m}+B_0
\end{pmatrix},\\
\De_1^{-1}&=\begin{pmatrix}
-\frac1m\vek\pi\cdot\vek A-A_0 & \frac{\im\Delta}{2\kf}(-A_++A_+^T),\\
-\frac{\im\Delta}{2\kf}(-A_-+A_-^T) & -\frac1m\tilde{\vek\pi}\cdot\vek A^T+A_0^T
\end{pmatrix},\\
\notag
\De_2^{-1}&=\begin{pmatrix}
\frac{\vek A^2}{2m} & 0\\
0 & -\frac{(\vek A^T)^2}{2m}
\end{pmatrix},
\end{align}
where we introduced the notation $\vek\pi\equiv\vek p-\vek B$, $\tilde{\vek\pi}\equiv\vek p+\vek B$, $p_\pm\equiv p_x\pm\im p_y$, and similarly for other quantities. The zeroth, first and second-order piece of the action in the expansion in the $\gr{SU(2)_S}$ gauge field now read
\begin{align}
\label{Seff}
-S_\text{eff}={}&\frac12\Tr\log\De_0^{-1}+\frac12\Tr(\De_0\De_1^{-1})\\
\notag
&+\frac14\Tr(2\De_0\De_2^{-1}-\De_0\De_1^{-1}\De_0\De_1^{-1})+\dotsb.
\end{align}
The propagator $\De_0$ is obtained by inverting the BdG operator $\De_0^{-1}$ and in momentum space takes the form
\begin{align}
\De_0&=\frac1\det\begin{pmatrix}
\im\omega-\frac{\tilde{\vek\pi}^2}{2m}+B_0 & -\frac{\im\Delta p_+}{\kf}\\
+\frac{\im\Delta p_-}{\kf} & \im\omega+\frac{\vek\pi^2}{2m}-B_0
\end{pmatrix},\\
\notag
\det&\equiv-\left(\omega+\frac{\im\vek p\cdot\vek B}m\right)^2-\left(\frac{\vek p^2+\vek B^2}{2m}-B_0\right)^2-\frac{\Delta^2\vek p^2}{\kf^2}.
\end{align}
As a consistency check, note that the last expression implies that for $\vek B=\vek0$ and $B_0=\mu$, the well-known spectrum of fermion excitations in the mean-field approximation follows,
\begin{equation}
E(\vek p)=\sqrt{\left(\frac{\vek p^2}{2m}-\mu\right)^2+\frac{\Delta^2\vek p^2}{\kf^2}}.
\end{equation}

The leading-order, pure superfluid part of the effective action is given by the first term in Eq.~(\ref{Seff}). The corresponding effective Lagrangian reads
\begin{align}
\label{LOlag}
\La_\text{eff}^\text{LO}={}&-\int\frac{\dd\omega\,\dd^2\vek p}{(2\pi)^3}\\
\notag
&\times\log\left[\omega^2+\left(\frac{\vek p^2+\vek B^2}{2m}-B_0\right)^2+\frac{\Delta^2\vek p^2}{\kf^2}\right],
\end{align}
and upon frequency integration,
\begin{equation}
\La_\text{eff}^\text{LO}=-\int\frac{\dd^2\vek p}{(2\pi)^2}\sqrt{\left(\frac{\vek p^2+\vek B^2}{2m}-B_0\right)^2+\frac{\Delta^2\vek p^2}{\kf^2}}.
\end{equation}
The effective Lagrangian is a (nonlinear) function of the combination $\frac{\vek B^2}{2m}-B_0$, as dictated by Galilei invariance.

The next-to-leading order of the effective action is given by the term quadratic in the $\gr{SU(2)_S}$ gauge field $A_\mu$. A straightforward, if slightly tedious, manipulation leads to the following expression,
\begin{widetext}
\begin{align}
\label{NLOlag}
\La_\text{eff}^\text{NLO}={}\frac14\int\frac{\dd\omega\,\dd^2\vek p}{(2\pi)^3}\Biggl\{&\frac2\det\frac\beta m\langle\vek A\cdot\vek A\rangle+\frac1{\det^2}\biggl[2(\alpha^2+\beta^2)\biggl\langle\left(\frac{\vek p\cdot\vek A}m\right)^2+\left(A_0-\frac{\vek B\cdot\vek A}m\right)^2\biggr\rangle+\frac{8\Delta^2}{\kf^2}\frac\beta m\langle(\vek p\cdot\vek A_2)^2\rangle\\
\notag
&+\frac{2\Delta^2}{\kf^2}(\alpha^2-\beta^2)\langle\vek A_2\cdot\vek A_2\rangle+2\gamma^2\biggl\langle\frac{(\vek p\cdot\vek A)(\vek p\cdot\vek A)^T}{m^2}-\left(A_0-\frac{\vek B\cdot\vek A}m\right)\left(A_0-\frac{\vek B\cdot\vek A}m\right)^T\biggr\rangle\biggr]\Biggr\},
\end{align}
\end{widetext}
where the brackets $\langle\cdot\rangle$ indicate trace over the spin space, and we introduced the shorthand notation
\begin{equation}
\alpha\equiv\im\omega-\frac{\vek p\cdot\vek B}m,\quad
\beta\equiv\frac{\vek p^2+\vek B^2}{2m}-B_0,\quad
\gamma\equiv\frac{\Delta|\vek p|}{\kf}.
\end{equation}
In Eq.~(\ref{NLOlag}), $\vek A$ denotes the spatial part of the matrix-valued gauge field $A_\mu$, whereas $\vek A_2$ corresponds to its second spin component, i.e.~is also a \emph{matrix}. This notation makes the result independent of the choice of normalization of the $\gr{SU(2)_S}$ generators.

The frequency integration can easily be carried out analytically. The momentum integration is, however, potentially ultraviolet divergent and thus requires regularization. Here we will use dimensional regularization, modifying the integration region into a Euclidean space of dimension $d\equiv2-2\epsilon$. Upon some manipulation, it can be shown that the second spin component of $A_\mu$ drops out of the action. (One arrives at the same conclusion if regularization with a hard cutoff $\Lambda$ is used instead and the limit $\Lambda\to\infty$ is taken.) Denoting the remaining matrix-valued components as $A_{\perp \mu}=(A_{\perp0}, \vek A_\perp)$, the effective Lagrangian takes the form
\begin{equation}
\La_\text{eff}^\text{NLO}=\frac12c_1\langle\vek A_\perp\cdot\vek A_\perp\rangle+\frac12c_2\biggl\langle\left(A_{\perp0}-\frac{\vek B\cdot\vek A_\perp}m\right)^2\biggr\rangle.
\end{equation}
The coefficients $c_{1,2}$ can be read off Eq.~(\ref{NLOlag}). Upon frequency integration, they can be cast as
\begin{equation}
\begin{aligned}
c_1&=\frac1{2m}\int\frac{\dd^d\vek p}{(2\pi)^d}\frac{\gamma^2}{\sqrt{\beta^2+\gamma^2}(\sqrt{\beta^2+\gamma^2}+\beta)},\\
c_2&=-\frac12\int\frac{\dd^d\vek p}{(2\pi)^d}\frac{\gamma^2}{(\beta^2+\gamma^2)^{3/2}}.
\end{aligned}
\end{equation}
The coefficient $c_2$ is well-defined through a convergent integral. The coefficient $c_1$, on the other hand, is given by a logarithmically divergent integral. To estimate such an integral in practice requires the knowledge of the ultraviolet and infrared momentum scales, where the integration is effectively cut off. In the present problem, the inverse size of the hard core of the interatomic potential can be taken as the ultraviolet cutoff, whereas the inverse of the size of the sample provides an infrared cutoff.

We are now in a position to restore the dependence of the effective action on the collective fields $\hdb$ and $\theta$.  Using Eqs.~(\ref{gaugetransfo}), (\ref{groundstate}) and (\ref{U12}), it is straightforward to show that
\begin{equation}
\langle\vek A_\perp\cdot\vek A_\perp\rangle=\frac12(D_r\hdb)^2,
\end{equation}
where the covariant derivative in the vector notation is given by $D_\mu\hdb\equiv\de_\mu\hdb+\vek A_\mu\times\hdb$. Likewise, it readily follows upon analytical continuation to real time that
\begin{multline}
\biggl\langle\left(A_{\perp0}-\frac{\vek B\cdot\vek A_\perp}m\right)^2\biggr\rangle\\
=\frac12\left[D_t\hdb+\frac1m(\de_r\theta-B_r)D_r\hdb\right]^2.
\end{multline}
In the above expressions, $A_\mu$ and $B_\mu$ are not composite anymore, but rather denote the original external gauge fields of the $\gr{SU(2)_S\times U(1)_\phi}$ group.

We have thus recovered the effective spin Lagrangian density from Eq.~(\ref{lag}) (without the dipole term $\La_\text{dip}$). The phase velocity of the spin waves is determined by the parameters of the microscopic theory through
\begin{equation}
\cs^2=-\frac{c_1}{c_2}.
\end{equation}


\section{Galilei invariance of the helical texture}
\label{app:galilei}

Since we are discussing a superfluid system that does not require an underlying crystal lattice or substrate, the microscopic physics must be Galilei-invariant. One can thus ask the following question: how can we deduce the existence of the helical spin texture in the ground state in a reference frame where the background superflow vanishes?

First, the fact that the magnetic field is introduced through the temporal component of the $\gr{SU(2)_S}$ gauge field implies that we have to use an unusual, so-called electric, limit of electromagnetism~\cite{LeBellac1973} if we want the coupling to the background fields to maintain Galilei invariance. In this limit, the Maxwell equations miss the term that induces the Faraday effect (electromagnetic induction). The electromagnetic potentials $\varphi$ and $\vek A$ transform under a Galilei boost with velocity $\vek v$ as
\begin{equation}
\varphi'=\varphi,\qquad
\vek A'=\vek A-\epsilon_0\mu_0\varphi\vek v.
\end{equation}
Accordingly, the electric and magnetic fields $\vek E$ and $\vek B$ transform as
\begin{equation}
\vek E'=\vek E,\qquad
\vek B'=\vek B-\epsilon_0\mu_0\vek v\times\vek E.
\end{equation}
The combination of a constant magnetic field and zero electric field, imposed on our system, is therefore invariant under the Galilei transformations in this limit.

Second, equilibrium properties of a many-body system are generally described by a density matrix that follows from the principle of maximum entropy. The principle in turn dictates that we have to correctly take into account all macroscopic constraints on the state of the system. In a system with macroscopic motion such as the background superflow, this means that we need to introduce a Lagrange multiplier for the momentum operator.

To carry out this procedure properly, we first have to rewrite the canonical Hamiltonian in terms of the canonical variables, that is, the field $\hdb$ and the associated canonical momentum,
\begin{equation}
\vek\pi\equiv\frac{\de\La}{\de(\de_t\hdb)}=\tilde D_t\hdb,
\end{equation}
which is itself invariant under Galilei boosts. The Hamiltonian, defined by Eq.~(\ref{ham2}), is then rewritten as
\begin{equation}
\label{Hsup}
\Ha=\frac12\vek\pi^2-\vek\pi\cdot(\gamma\vek H\times\hdb+u_r\de_r\hdb)+\frac{\cs^2}2(\de_r\hdb)^2-\frac12\Omega_\text{L}^2(\hlb\cdot\hdb)^2.
\end{equation}
Next, we introduce the Lagrange multiplier $w_r$ for the operator of momentum density $\mathscr{P}_r$, given by the standard Noether expression
\begin{equation}
\mathscr{P}_r=-\frac{\de\La}{\de(\de_t\hdb)}\cdot\de_r\hdb=-\vek\pi\cdot\de_r\hdb.
\end{equation}
The grandcanonical Hamiltonian $\Ha_w$ for the spin wave sector is then obtained from the canonical Hamiltonian~(\ref{Hsup}) by subtracting the term $w_r\mathscr{P}_r$,
\begin{equation}
\begin{split}
\Ha_w={}&\Ha-w_r\mathscr{P}_r\\
={}&\frac12\vek\pi^2-\vek\pi\cdot(\gamma\vek H\times\hdb+\tilde u_r\de_r\hdb)\\
&+\frac{\cs^2}2(\de_r\hdb)^2-\frac12\Omega_\text{L}^2(\hlb\cdot\hdb)^2,
\end{split}
\label{Hcan}
\end{equation}
where $\tilde{\vek u}\equiv\vek u-\vek w$. Unlike the Hamiltonian $\Ha$, the grandcanonical Hamiltonian $\Ha_w$ is invariant under the simultaneous Galilei transformation of the coordinates and fields, whose infinitesimal form reads
\begin{equation}
\begin{gathered}
\vek x'=\vek x+\vek vt,\qquad
\vek w'=\vek w+\vek v,\\
\theta'(\vek x')=\theta(\vek x)+m\vek v\cdot\vek x.
\end{gathered}
\end{equation}
The many-body ground state of the system, which is determined by the absolute minimum of (the spatial integral of) $\Ha_w$, is therefore independent of the choice of reference frame, as it should.

To proceed towards finding the ground state, all one has to do is to cast Eq.~(\ref{Hcan}) as
\begin{equation}
\begin{split}
\Ha_w={}&\frac12(\de_t\hdb+w_r\de_r\hdb)^2-\frac12(\gamma\vek H\times\hdb+\tilde u_r\de_r\hdb)^2\\
&+\frac{\cs^2}2(\de_r\hdb)^2-\frac12\Omega_\text{L}^2(\hlb\cdot\hdb)^2,
\end{split}
\label{hamw}
\end{equation}
and then follow the argument below Eq.~(\ref{ham2}). The spatial profile of the ground state is still given by Eq.~(\ref{larmor}) upon replacing the superflow velocity $\vek u$ with the Galilei-invariant combination $\tilde{\vek u}=\vek u-\vek w$. In an arbitrarily chosen reference frame, the texture also has a nontrivial temporal profile given by $\de_t\hdb=-w_r\de_r\hdb$. This is a necessary consequence of Galilei invariance and the spatial dependence of the static texture we found in Eq.~(\ref{larmor}).

The spin density and current are given by the effective Lagrangian, Eq.~(\ref{lag}), in any inertial reference frame. The general expression for the Noether current of the $\gr{SU(2)_S}$ spin symmetry following from this Lagrangian is given by Eq.~(\ref{spincurrent}). Inserting the texture found by minimization of the grandcanonical Hamiltonian~(\ref{hamw}), one finds the following spin density and current, respectively,
\begin{equation}
\vek j^0=\frac{\cs^2}{\cs^2-\tilde u^2}\gamma\vek H,\qquad
\vek j_r=w_r\vek j^0.
\end{equation}
It follows that the spin density carried by the helical texture is Galilei-invariant, whereas the spin current transforms as a Galilei vector. The Lagrange multiplier $\vek w$ plays the role of the velocity of the spin degrees of freedom.

We conclude that the helical texture, discovered in Sec.~\ref{sec:groundstate} in the frame where $\vek w=\vek0$, can be obtained as well for instance in the frame where there is no background superflow. All that matters is the \emph{relative} motion of the superfluid and spin degrees of freedom, encoded by their relative velocity $\tilde{\vek u}$.


\bibliography{library}
\bibliographystyle{apsrev4-1}


\end{document}